\documentclass[twocolumn,letterpaper]{IEEEAerospaceCLS}

\usepackage{subcaption}
\usepackage{graphicx}
\usepackage{amsmath}
\usepackage{tikz}
\usepackage{physics}
\usepackage{amssymb}
\usepackage{bm}
\usepackage{algorithm}
\usepackage{algpseudocode}
\usepackage[pdftex]{hyperref}
\usepackage{mathtools}
\usepackage{tabularx}
\usepackage[normalem]{ulem}

\newcommand{\stkout}[1]{\ifmmode\text{\sout{\ensuremath{#1}}}\else\sout{#1}\fi}

\DeclareUnicodeCharacter{2212}{-}

\begin{document}
\title{One-Shot Initial Orbit Determination in Low-Earth Orbit}

\author{%
Ricardo Ferreira\\ 
Department of Computer Science\\
NOVA School of Science and Technology\\
Caparica, Portugal\\
rjn.ferreira@campus.fct.unl.pt
\and
Marta Guimarães\\ 
AI Department\\
Neuraspace\\
Coimbra, Portugal\\
marta.guimaraes@neuraspace.com
\and
Filipa Valdeira\\
Department of Computer Science\\
NOVA School of Science and Technology\\
Caparica, Portugal\\
f.valdeira@campus.fct.unl.pt
\and
Cláudia Soares\\ 
Department of Computer Science\\
NOVA School of Science and Technology\\
Caparica, Portugal\\
claudia.soares@fct.unl.pt
%%%% IMPORTANT: Use the correct copyright information--IEEE, Crown, or U.S. government. %%%%%
\thanks{\footnotesize 979-8-3503-0462-6/24/$\$31.00$ \copyright2024 IEEE}              % This creates the copyright info that is the correct 2024 data.
%\thanks{{U.S. Government work not protected by U.S. copyright}}         % Use this copyright notice only if you are employed by the U.S. Government.
%\thanks{{979-8-3503-0462-6/24/$\$31.00$ \copyright2024 Crown}}          % Use this copyright notice only if you are employed by a crown government (e.g., Canada, UK, Australia).
%\thanks{{979-8-3503-0462-6/24/$\$31.00$ \copyright2024 European Union}}    % Use this copyright notice if you are employed by the European Union.
}

\maketitle

\thispagestyle{plain}
\pagestyle{plain}

\begin{abstract}
Due to the importance of satellites for society and the exponential increase in the number of objects in orbit, whether they are space debris or functional satellites, it is important to accurately determine the state (e.g., position and velocity) of these Resident Space Objects (RSOs) at any time and in a timely manner. State-of-the-art methodologies for initial orbit determination consist of Kalman-type filters that process sequential data over time and return the state and associated uncertainty of the object, as is the case of the Extended Kalman Filter (EKF). However, these methodologies are dependent on a good initial guess for the state vector and usually simplify the physical dynamical model, due to the difficulty of precisely modeling perturbative forces, such as atmospheric drag and solar radiation pressure. Other approaches do not require assumptions about the dynamical system, such as the trilateration method, and require simultaneous measurements, such as three measurements of range and range-rate for the particular case of trilateration. We consider the same setting of simultaneous measurements (one-shot), resorting to time delay and Doppler shift measurements. Based on recent advancements in the problem of moving target localization for sonar multistatic systems, we are able to formulate the problem of initial orbit determination as a Weighted Least Squares. With this approach, we are able to directly obtain the state of the object (position and velocity) and the associated covariance matrix from the Fisher's Information Matrix (FIM). We demonstrate that, for small noise, our estimator is able to attain the Cramér-Rao Lower Bound accuracy, i.e., the accuracy attained by the unbiased estimator with minimum variance. We also numerically demonstrate that our estimator is able to attain better accuracy on the state estimation than the trilateration method and returns a smaller uncertainty associated with the estimation.
\end{abstract}

\tableofcontents

\section{Introduction}

From the most recent data, it is estimated that 7500 functional satellites share their orbits with millions of space debris objects~\cite{esa-numbers}. Due to the important role of these satellites for society, it is important to accurately determine the state (e.g., position and velocity) of these Resident Space Objects (RSOs) at any time.

At this point in time, the Extended Kalman Filter (EKF) is the standard method for the initial orbit determination process~\cite{schutz2004statistical,vallado2001fundamentals,lam2010analysis,pardal2011robustness}. This approach has revealed useful for processing sequential data over time, returning the state vector of the RSO, i.e., position and velocity of the object at an initial instant, as well as the covariance matrix associated with the estimation. 

However, the performance of the EKF is dependent on a good initial guess for the initial state vector, which will serve to define the reference trajectory. With this reference trajectory, through a first-order Taylor Expansion at each observable instant, we can formulate the difference between our reference trajectory and the true trajectory as a linear system, and then apply the standard Kalman filter, which iteratively converges to the true state vector, when the right initial conditions are chosen~\cite{krener2003convergence}. This approach is particularly useful when an orbital state is available (usually for cataloged objects that are constantly monitored) and new observations can be used in a filtering approach, such as the Extended Kalman Filter (EKF)

To obtain a reference trajectory, we need to propagate the initial conditions forward in time. Orbit propagation is difficult to model precisely, especially in orbital mechanics, due to the existence of perturbative forces (such as atmospheric drag and solar radiation pressure)~\cite{vallado2001fundamentals}. Typically, simplifications of the physical dynamical model are adopted, which translates into a structural uncertainty that is not taken into account~\cite{poore2016covariance}. Also, orbit propagation is frequently associated with the integration of Ordinary Differential Equations (ODEs), whose processing time depends on the desired precision.

Other variants of the Kalman filter, such as the Unscented Kalman Filter (UKF)~\cite{julier1997new,pardal2011robustness} or the Second-Order Extended Kalman Filter~\cite{einicke2012smoothing}, improve the performance of EKF by modeling more precisely the nonlinear dynamics. However, they remain dependent on the propagation of a reference trajectory, whose drawbacks were already listed.

Nevertheless, the problem of initial orbit determination becomes especially relevant in cases where we observe an object for the first time, with no prior information about the object's orbital state.

Over the years, different approaches have been presented to the problem of initial orbit determination of near-Earth orbiting satellites, when there is no \textit{a priori} knowledge about the orbit of the object. Some of these approaches only consider angle observations to obtain three position vectors at different instants, such as Gauss's method and Double-r iteration~\cite{vallado2001fundamentals}. The accuracy of Gauss's method is very sensitive to the separation between observations, preferably less than 10 degrees. Double-r iteration, incorporated in a method proposed by Escobal~\cite{escobal1970methods} is able to handle observations that are days apart. To obtain the full-state vector, one can resort to Gibbs method~\cite{gibbs1889determination}, which performs best for larger time lengths between position vectors, or the Herrick-Gibbs method~\cite{herrick1971astrodynamics}, which was developed for smaller time lengths when the vectors are almost parallel~\cite{kaushik2016statistical}. These methods, given three position vectors, are able to obtain the velocity vector at the middle point. Another solution to obtain the velocity vector is to solve Lambert's problem~\cite{gooding1996new,izzo2015revisiting,gooding1988solution,gooding1990procedure,lancaster1969unified}, which determines the Keplerian elements given two position vector and the time period between the two positions, often called \textit{time of flight}. A disadvantage of these methods is the fact that they assume that the orbit is not perturbed, which for Low-Earth Orbit satellites is not true due to the impact of previously mentioned perturbative forces.

Other approaches consider the setting where a small number of observations are available (\textit{very-short-arcs})~\cite{shang2019VSA,qu2022VSA}. State-of-the-art approaches gather new information from high-order kinematic parameters which can be obtained from time derivatives of radar's echo phase, however, the authors state that this is only possible for Low-Earth objects with stable attitude.

In this work, we consider the scenario where all measurements used for the estimation are taken from the same instant (one-shot). There are other approaches that consider the same setting. The \emph{trilateration} method~\cite{escobal1970methods,vallado2001fundamentals,hough2012precise}, does not need to formulate assumptions about the dynamical systems, and is able to accurately determine the state vector and associated uncertainty, requiring, simultaneously, three range measurements and three range-rate measurements~\cite{hough2012precise}. We consider that the radars are able to gather time delay and Doppler shift measurements.

Our goal is to develop an estimator for the problem of initial orbit determination for Low-Earth Orbit (LEO). Unlike the EKF, our method does not require the propagation of a reference trajectory. Consequently, it does not rely on assumptions about the physical dynamical system and does not require linear approximations.

Based on recent advancements, we adapt the formulation for the problem of moving target localization for sonar multistatic systems~\cite{rui2015efficient,yang2016moving,einemo2015weighted} in the two-dimensional space, to the problem of initial orbit determination, in the three-dimensional plane. 

Following the work presented by Yang et al.~\cite{yang2016moving}, we estimate the state vector in two steps. First, we formulate the problem as a Weighted Least Squares (WLS) through the use of intermediate variables. Then, we correct our estimate by relating the intermediate variables and the variable for the position and velocity of the object. From this formulation, it is possible to directly obtain the state of the object (position and velocity) as well as the covariance matrix from the Fisher's Information Matrix (FIM)~\cite{bishop2006pattern}. Similar to~\cite{yang2016moving}, we show that in the context of space applications, the method is able to attain the Cramér-Rao Lower Bound (CRLB) accuracy for small noise, i.e., the accuracy attained by the unbiased estimator with minimum variance~\cite{kay1993fundamentals}.

\subsection{Contributions}

Compared with the state-of-the-art approaches for the problem of initial orbit determination, our approach retains the following advantages and differences:

\begin{itemize}
    \item Our method does not need to propagate a reference trajectory, thus avoiding linear approximations and simplifications of the nonlinear dynamical systems;
    \item We present a non-iterative estimator for the problem of initial orbit determination, providing the complete state vector of the object (position and velocity) and the corresponding associated covariance matrix;
    \item Due to the formulation of the problem as a Weighted Least Squares (WLS), we can easily obtain the solution in closed form, therefore presenting constant computational complexity;
    \item Similarly to the results for the sonar multistatic system, we demonstrate that, for small noise, the solution is able to attain CRLB accuracy in the context of space application.
\end{itemize}

\section{Problem Formulation}

In this section, we formulate the problem of locating an object using time delay and Doppler shift measurements. Then, we present a two-staged localization algorithm by formulating the problem as Weighted Least Squares (WLS) and obtaining the solution in closed form. 

Consider a multistatic radar system composed by $M$ transmitters located at ${\mathbf{t}}_i \in {\mathbb{R}}^{3}$, for $i = 1, \ldots, M$, and $N$ receivers located at ${\mathbf{s}}_j \in {\mathbb{R}}^{3}$, for $j = 1, \ldots, N$. The goal is to locate an RSO, whose position and velocity are denoted by ${\mathbf{x}} \in {\mathbb{R}}^{3}$ and ${\mathbf{v}} \in {\mathbb{R}}^{3}$, respectively. 

We assume that is possible to obtain two measurements from each pair of transmitter and receiver: time delay and Doppler shift measurements. The true time delay, $\tau_{ij}^0$, which is the time that the signal takes from the transmitter $i$ to the target and then to the receiver $s$, can be modeled as

\begin{equation}
    \label{eq:differential-delay-model}
    \centering
    \begin{aligned}
    \tau_{ij}^0 = \frac{\norm{\mathbf{x} - \mathbf{t}_i} + \norm{\mathbf{x} - \mathbf{s}_j}}{c},
    \end{aligned}
\end{equation}
where $c$ denotes the speed of light and $\norm{\cdot}$ denotes the Euclidean norm. So, the time delay measurement is given by

\begin{equation}
    \label{eq:differential-delay-measurement_rearranged}
    \centering
    \begin{aligned}
    \tau_{ij} = \tau_{ij}^0 + \Delta \tau_{ij},
    \end{aligned}
\end{equation}
where $\Delta \tau_{ij}$ denotes the time delay measurement noise, which we assume follows a Gaussian distribution with mean zero and standard deviation $\sigma_{\tau}$. We can collect all the measurements in one vector as

\begin{equation}
    \label{eq:differential-delay-measurement_vector}
    \centering
    \begin{aligned}
    \bm{\tau} = \begin{bmatrix}
        \bm{\tau}_1 \\
        \vdots \\
        \bm{\tau}_M
    \end{bmatrix} = \bm{\tau}^0 + \Delta \bm{\tau},
    \end{aligned}
\end{equation}
such that $\bm{\tau}_i \coloneqq \left(\tau_{i1}, \ldots, \tau_{iN}\right)$ denotes the measurements between the transmitter $i$ and the $N$ receivers.

The true Doppler shift, $f_{ij}^0$, which results from the motion of the satellite, can be modeled as

\begin{equation}
    \label{eq:doppler-shift-model}
    \centering
    \begin{aligned}
    f_{ij}^0 = \frac{f_{c,i}}{c} \qty( \rho_{\mathbf{x}, \mathbf{t}_i}^T \mathbf{v} + \rho_{\mathbf{x}, \mathbf{s}_j}^T \mathbf{v} ),
    \end{aligned}
\end{equation}
such that $f_{c,i}$ denotes the carrier frequency of the signal from transmitter $i$ and 

\begin{equation}
    \centering
    \begin{aligned}
    \rho_{\mathbf{x}, \mathbf{t}_i} = \frac{\mathbf{x} - \mathbf{t}_i}{\norm{\mathbf{x} - \mathbf{t}_i}}, \quad \rho_{\mathbf{x}, \mathbf{s}_j} = \frac{\mathbf{x} - \mathbf{s}_j}{\norm{\mathbf{x} - \mathbf{s}_j}}.
    \end{aligned}
\end{equation}
Similarly, the Doppler shift measurement is given by

\begin{equation}
    \label{eq:doppler-shift-measurement}
    \centering
    \begin{aligned}
    f_{ij} = f_{ij}^0 + \Delta f_{ij},
    \end{aligned}
\end{equation}
where $\Delta f_{ij}$ denotes the Doppler shift measurement noise, which we also assume follows a Gaussian distribution with mean zero and standard deviation $\sigma_{f}$. Analogously, we can collect all the Doppler shift measurements in one vector as

\begin{equation}
    \centering
    \begin{aligned}
    \bm{f} = \begin{bmatrix}
        \bm{f}_1 \\
        \vdots \\
        \bm{f}_M
    \end{bmatrix} = \bm{f}^0 + \Delta \bm{f},
    \end{aligned}
\end{equation}
such that $\bm{f}_i \coloneqq \left(\bm{f}_{i1}, \ldots, \bm{f}_{iN}\right)$ denotes the measurements between the transmitter $i$ and the $N$ receivers. 

\subsection{Localization Algorithm - Estimation Step}

By rearranging~\eqref{eq:differential-delay-model} as

\begin{equation}
    \label{eq:differential-delay-relation}
    \centering
    \begin{aligned}
    c \tau_{ij}^0 - \norm{\mathbf{x} - \mathbf{t}_i} = \norm{\mathbf{x} - \mathbf{s}_j},
    \end{aligned}
\end{equation}
substituting $\tau_{ij}^0 = \tau_{ij} - \Delta \tau_{ij}$ and squaring both sides of the equation, after some manipulation, we obtain

\begin{equation}
    \label{eq:differential-delay-noise-relaxation}
    \centering
    \begin{aligned}
    2 c \Delta \tau_{ij} \norm{\mathbf{x} - \mathbf{s}_j} &\approx \ c^2 \tau_{ij}^2 + \norm{\mathbf{t}_i}^2 - \norm{\mathbf{s}_j}^2 \\ 
    &- 2 \qty(\mathbf{t}_i - \mathbf{s}_j)^T {\mathbf{x}} - 2 c \tau_{ij} \norm{\mathbf{x} - \mathbf{t}_i}.
    \end{aligned}
\end{equation}
Similarly, by manipulating the expression of the true Doppler shift~\eqref{eq:doppler-shift-model}, after some manipulation, we arrive at the expression 

\begin{equation}
    \label{eq:doppler-shift-noise-relation}
    \centering
    \begin{aligned}
    &2 c f_{c,i} \rho_{\mathbf{x}, \mathbf{s}_j}^T {\mathbf{v}} \Delta \tau_{ij} + 2 c \norm{\mathbf{x} - \mathbf{s}_j} \Delta f_{ij} \approx 2 c^2 \tau_{ij} f_{ij} \\
    &+ 2 f_{c,i} \qty(\mathbf{t}_i - \mathbf{s}_j)^T {\mathbf{v}} - 2 c f_{ij} \norm{\mathbf{x} - \mathbf{t}_i} - 2 c f_{c,i} \tau_{ij} \rho_{\mathbf{x}, \mathbf{t}_i}^T {\mathbf{v}}.
    \end{aligned}
\end{equation}
In both~\eqref{eq:differential-delay-noise-relaxation} and~\eqref{eq:doppler-shift-noise-relation} the second-order noise terms are ignored. In Appendix~\ref{app:stage1-noise-terms-relation}, we present a detailed derivation of these expressions.

To overcome the problem of non-linearity with respect to the variables $\mathbf{x}$ and $\mathbf{v}$ due to the terms $\norm{\mathbf{x} - \mathbf{t}_i}$ and $\rho_{\mathbf{x}, \mathbf{t}_i}^T \mathbf{v}$, for $i = 1, \ldots, M$, we can express the unknown variable, $\mathbf{y}$, as

\begin{equation}
    \centering
    \begin{aligned}
    \mathbf{y} = \begin{bmatrix}
        \mathbf{x} \\
        \mathbf{v} \\
        \norm{\mathbf{x} - \mathbf{t}_1} \\
        \vdots \\
        \norm{\mathbf{x} - \mathbf{t}_M} \\
        \rho_{\mathbf{x}, \mathbf{t}_1}^T \mathbf{v} \\
        \vdots \\
        \rho_{\mathbf{x}, \mathbf{t}_M}^T \mathbf{v}
    \end{bmatrix}.
    \end{aligned}
\end{equation}
In this way, we can express ~\eqref{eq:differential-delay-noise-relaxation} and~\eqref{eq:doppler-shift-noise-relation} as

\begin{equation}
    \label{eq:compact-noise-relation}
    \centering
    \begin{aligned}
    B_{\tau} \Delta \bm{\tau} &\approx b_{\tau} - A_{\tau} \mathbf{y} \\
    B_{f_1} \Delta \bm{\tau} + B_{f_2} \Delta \bm{f} &\approx b_{f} - A_{f} \mathbf{y},
    \end{aligned}
\end{equation}
where

\begin{equation}
    \centering
    \begin{aligned}
    \qty(b_{\tau})_{N (i - 1) + j} &= c^2 \tau_{ij}^2 + \norm{\mathbf{t}_i}^2 - \norm{\mathbf{s}_j}^2 \\
    \qty(b_{f})_{N (i - 1) + j} &= 2 c^2 \tau_{ij} f_{ij},
    \end{aligned}
\end{equation}
for $i = 1, \ldots, M$ and $j = 1, \ldots, N$, such that $(p)_{i}$ denotes the $i$-th entry of the vector $p$. The matrices $A_{\tau}$ and $A_{f}$ can be expressed as

\begin{equation}
    \centering
    \begin{aligned}
    \qty(A_{\tau})_{N (i - 1) + j} &= \begin{bmatrix}
        2 \qty(\mathbf{t}_i - \mathbf{s}_j) \\  {\mathbf{0}}_{\qty(3 + i - 1)} \\ 2 c \tau_{ij} \\ {\mathbf{0}}_{\qty(2M - i)}
    \end{bmatrix}^T \\
    \qty(A_{f})_{N (i - 1) + j} &= \begin{bmatrix}
        {\mathbf{0}}_{3} \\ 2 f_{c,i} \qty(\mathbf{t}_i - \mathbf{s}_j) \\  {\mathbf{0}}_{\qty(i - 1)} \\ 2 c f_{ij} \\ {\mathbf{0}}_{M} \\ 2 c f_{c,i} \tau_{ij} \\ {\mathbf{0}}_{\qty(M - i)}
    \end{bmatrix}^T ,
    \end{aligned}
\end{equation}
for $i = 1, \ldots, M$ and $j = 1, \ldots, N$, such that ${\mathbf{0}}_{p} \in {\mathbb{R}}^{p}$ denotes a vector whose all entries are zero and $(D)_{i}$ denotes the $i$-th row of matrix $D$. The matrices $B_{\tau}$, $B_{f_1}$ and $B_{f_2}$ can be expressed as $B_{\tau} = \mathbf{I}_M \otimes \mbox{diag}\qty(\mathbf{r^0})$, $B_{f_1} = \mbox{diag}\qty(\bm{f_c}) \otimes \mbox{diag}\qty(\mathbf{\dot{r}^0})$ and $B_{f_2} = \mathbf{I}_M \otimes \mbox{diag}\qty(\mathbf{r^0})$, where $\mathbf{I}_p$ denotes the $p \times p$ identity matrix, $\otimes$ denotes the Kronecker product and $\mbox{diag}\qty(\mathbf{r})$ denotes the diagonal matrix whose diagonal is the vector $\mathbf{r}$, with

\begin{equation}
    \centering
    \begin{aligned}
    &\bm{f_c} = \begin{bmatrix}
        f_{c,1} & \cdots & f_{c,M}
    \end{bmatrix}^T, \\
    &\mathbf{r^0} = \begin{bmatrix}
        \norm{\mathbf{x} - \mathbf{s}_1} & \cdots & \norm{\mathbf{x} - \mathbf{s}_N}
    \end{bmatrix}^T, \\
    &\mathbf{\dot{r}^0} = \begin{bmatrix}
        \rho_{\mathbf{x}, \mathbf{s}_1}^T \mathbf{v} & \cdots & \rho_{\mathbf{x}, \mathbf{s}_N}^T \mathbf{v}
    \end{bmatrix}^T.
    \end{aligned}
\end{equation}
By stacking the expressions in~\eqref{eq:compact-noise-relation} as

\begin{equation}
    \label{eq:compact-stage1-noise-relation}
    \centering
    \begin{aligned}
    & B \Delta \bm{\alpha} \approx b - A \mathbf{y}
    \end{aligned},
\end{equation}
such that $b = \begin{bmatrix}
    b_{\tau}^T & b_{f}^T 
\end{bmatrix}^T$, $A = \begin{bmatrix}
    A_{\tau}^T & A_{f}^T 
\end{bmatrix}^T$, $\Delta \bm{\alpha} = \begin{bmatrix}\Delta \bm{\tau}^T & \Delta \bm{f}^T \end{bmatrix}^T$ and 

\begin{equation}
    \centering
    \begin{aligned}
    B = 2 c \, \begin{bmatrix}
        B_{\tau} & {\mathbf{0}}_{MN \times MN} \\
        B_{f_1} & B_{f_2}
    \end{bmatrix},
    \end{aligned}
\end{equation}
where ${\mathbf{0}}_{p \times p} \in {\mathbb{R}}^{p \times p}$ denotes a matrix whose entries are all zero, we can define the target localization problem as a Weighted Least Squares (WLS) as

\begin{equation}
    \centering
    \begin{aligned}
    \underset{\mathbf{y}}{\mbox{minimize}} \quad & \left(b - A \mathbf{y}\right)^T W_{\alpha} \left(b - A \mathbf{y}\right),
    \end{aligned}
\end{equation}
such that $W_{\alpha} = \qty(B Q_{\alpha} B^T)^{-1}$, whose solution is given by~\cite{bishop2006pattern}

\begin{equation}
    \centering
    \begin{aligned}
    \Tilde{\mathbf{y}} = \qty(A^T W_{\alpha} A)^{-1} A^T W_{\alpha} b .
    \end{aligned}
\end{equation}
The weighting matrix, $W_{\alpha}$, is dependent on the position and velocity of the object through $B$. To overcome this problem, we can set $W_{\alpha} = Q_{\alpha}^{-1}$ to obtain an initial guess for $\mathbf{y}$, and then use the initial guess for $\mathbf{x}$ and $\mathbf{v}$ from the unknown vector to construct the matrix $B$, obtaining a final estimation for the first stage of the localization algorithm.

\subsection{Localization Algorithm - Correction Step}

In the first stage of the localization algorithm, the relation between the intermediate variables $\gamma_i = \norm{\mathbf{x} - \mathbf{t}_i}$ and $\beta_i = \rho_{\mathbf{x}, \mathbf{t}_i}^T \mathbf{v}$, for $i = 1, \ldots, M$, and the variables for the position and velocity of the object, $\mathbf{x}$ and $\mathbf{v}$, is neglected.

Let $\mathbf{\Tilde{y}}$ denote the solution obtained from the first stage, which for small noise can be considered to have negligible bias~\cite{yang2016moving}, whose covariance matrix is $\mbox{cov}\qty(\mathbf{\Tilde{y}}) = \qty(A^T W_{\alpha} A)^{-1}$. We can formulate $\Tilde{\mathbf{y}} = {\mathbf{y}} + \Delta \Tilde{\mathbf{y}}$, such that $\mathbf{y}$ denotes the unknown variables and $\Delta \mathbf{\Tilde{y}}$ the first stage estimation error, where

\begin{equation}
    \centering
    \begin{aligned}
    \Delta \mathbf{\Tilde{y}} = \begin{bmatrix}
        \Delta \Tilde{\mathbf{x}} \\
        \Delta \Tilde{\mathbf{v}} \\
        \Delta \Tilde{\gamma}_1 \\
        \vdots \\
        \Delta \Tilde{\gamma}_M \\
        \Delta \Tilde{\beta}_1 \\
        \vdots \\
        \Delta \Tilde{\beta}_M
    \end{bmatrix}.
    \end{aligned}
\end{equation}
We can approximately relate the intermediate variables with the position and velocity unknown variables as

\begin{subequations}
\begin{equation}
    \label{eq:stage2-gamma-intermediate-noise-terms-relation}
    \centering
    \begin{aligned}
        2 \Tilde{\gamma}_i \Delta \Tilde{\gamma}_i \approx \Tilde{\gamma}_i^2 &- \Tilde{\mathbf{x}}^T \Tilde{\mathbf{x}} + 2 {\mathbf{t}}_i^T \Tilde{\mathbf{x}} - {\mathbf{t}}_i^T {\mathbf{t}}_i \\
        &+ 2 \qty(\Tilde{\mathbf{x}} - {\mathbf{t}}_i)^T \Delta {\mathbf{x}}
    \end{aligned}
\end{equation}
\begin{equation}
    \label{eq:stage2-beta-intermediate-noise-terms-relation}
    \centering
    \begin{aligned}
        \Tilde{\beta}_i \Delta \Tilde{\gamma}_i + \Tilde{\gamma}_i \Delta \Tilde{\beta}_i &\approx \Tilde{\gamma}_i \Tilde{\beta}_i - \Tilde{\mathbf{x}}^T \Tilde{\mathbf{v}} + {\mathbf{t}}_i^T \Tilde{\mathbf{v}} \\ 
        &+ \Tilde{\mathbf{v}}^T \Delta {\mathbf{x}} + \qty(\Tilde{\mathbf{x}} - {\mathbf{t}}_i)^T \Delta {\mathbf{v}},
    \end{aligned}
\end{equation}
\end{subequations}
where $\Delta \mathbf{x}$ and $\Delta \mathbf{v}$ are random variables denoting the true error between the true position and velocity, $\mathbf{x}$ and $\mathbf{v}$, and the first stage estimates, $\Tilde{\mathbf{x}}$ and $\Tilde{\mathbf{v}}$, respectively. 
As in the first stage of the localization algorithm, the second-order noise terms are ignored, whereas similarly, for small levels of noise, the introduced bias can be neglected. In Appendix~\ref{app:stage2-noise-terms-relation}, we present a detailed derivation of these expressions. In order to construct the estimator from all the values in $\Delta \mathbf{\Tilde{y}}$, we introduce the equations

\begin{equation}
    \label{eq:stage2-noise-terms-relation}
    \centering
    \begin{aligned}
    \Delta \Tilde{\mathbf{x}} = \Delta \mathbf{x} \\
    \Delta \Tilde{\mathbf{v}} = \Delta \mathbf{v}
    \end{aligned}
\end{equation}
that expresses the relation between the random variables for the position and velocity noise terms (in the right side of the equation) with the position and velocity random noise errors (in the left side of the equation), where we assume that the noise random have zero mean since the bias from the first stage estimation is negligible.

For the second stage of the localization algorithm, we consider that the problem's variables are $\Delta \mathbf{x}$ and $\Delta \mathbf{v}$, which will translate into the amount of correction to the estimate from the first stage.

By defining the second stage random variable as 

\begin{equation}
    \centering
    \begin{aligned}
    \mathbf{z} = \begin{bmatrix}
        \Delta \mathbf{x} \\
        \Delta \mathbf{v}
    \end{bmatrix},
    \end{aligned}
\end{equation}
we can stack~\eqref{eq:stage2-gamma-intermediate-noise-terms-relation},~\eqref{eq:stage2-beta-intermediate-noise-terms-relation} and~\eqref{eq:stage2-noise-terms-relation} as

\begin{equation}
    \label{eq:compact-stage2-noise-relation}
    \centering
    \begin{aligned}
    & B_2 \Delta \Tilde{\mathbf{y}} \approx h - G {\mathbf{z}},
    \end{aligned}
\end{equation}
such that $h = \begin{bmatrix} h_{\gamma}^T & h_{\beta}^T  & {{\mathbf{0}}}_6^T \end{bmatrix}^T$, where

\begin{equation}
    \centering
    \begin{aligned}
    \qty(h_{\gamma})_i &= \Tilde{\gamma}_i^2 - \Tilde{\mathbf{x}}^T \Tilde{\mathbf{x}} + 2 {\mathbf{t}}_i^T \Tilde{\mathbf{x}} - {\mathbf{t}}_i^T {\mathbf{t}}_i \\
    \qty(h_{\beta})_i &= \Tilde{\gamma}_i \Tilde{\beta}_i - \Tilde{\mathbf{x}}^T \mathbf{\Tilde{v}} + {\mathbf{t}}_i^T \Tilde{\mathbf{v}},
    \end{aligned}
\end{equation}
and $G = \begin{bmatrix}
    G_{\gamma}^T & G_{\beta}^T & -\mathbf{I}_{6}
\end{bmatrix}^T$ such that

\begin{equation}
    \centering
    \begin{aligned}
    \qty(G_{\gamma})_i &= \begin{bmatrix}
        -2 \qty(\Tilde{\mathbf{x}} - {\mathbf{t}}_i)^T & \ \ {{\mathbf{0}}}_{3}^T
    \end{bmatrix} \\
    \qty(G_{\beta})_i &= \begin{bmatrix}
        -\Tilde{\mathbf{v}} & \ \ - \qty(\Tilde{\mathbf{x}} - {\mathbf{t}}_i)^T
    \end{bmatrix}.
    \end{aligned}
\end{equation}
The matrix $B_2$ can be constructed as

\begin{equation}
    \centering
    \begin{aligned}
    B_2 = \begin{bmatrix}
        {\mathbf{0}}_{M \times 3} & {\mathbf{0}}_{M \times 3} & 2 \mbox{diag}\qty(\bm{\gamma}) & {\mathbf{0}}_{M \times M} \\
        {\mathbf{0}}_{M \times 3} & {\mathbf{0}}_{M \times 3} & \mbox{diag}\qty(\bm{\beta}) & \mbox{diag}\qty(\bm{\gamma}) \\
        {\mathbf{I}}_{3} & {\mathbf{0}}_{3 \times 3} & {\mathbf{0}}_{3 \times M} & {\mathbf{0}}_{3 \times M} \\
        {\mathbf{0}}_{3 \times 3} & {\mathbf{I}}_{3} & {\mathbf{0}}_{3 \times M} & {\mathbf{0}}_{3 \times M} \\
    \end{bmatrix}.
    \end{aligned}
\end{equation}

In this way, we can define the second stage of the target localization problem as a Weighted Least Squares (WLS) as

\begin{equation}
    \centering
    \begin{aligned}
    \underset{\mathbf{z}}{\mbox{minimize}} \quad & \left(h - G {\mathbf{z}}\right)^T W_{2} \left(h - G {\mathbf{z}}\right)
    \end{aligned}
\end{equation}
such that $W_{2} = \qty(B_2 \ \mbox{cov}\qty(\Tilde{\mathbf{y}}) \ B_2^T)^{-1}$, whose solution is given by

\begin{equation}
    \centering
    \begin{aligned}
    {\mathbf{z}} = \begin{bmatrix}
        \Delta \Bar{\mathbf{x}}^T & \Delta \Bar{\mathbf{v}}^T
    \end{bmatrix}^T = \qty(G^T W_{2} G)^{-1} G^T W_{2} h .
    \end{aligned}
\end{equation}
The final estimates are given by

\begin{equation}
    \centering
    \begin{aligned}
    \mathbf{x} &= \Tilde{\mathbf{x}} - \Delta \Bar{\mathbf{x}} \\
    \mathbf{v} &= \Tilde{\mathbf{v}} - \Delta \Bar{\mathbf{v}}
    \end{aligned}
\end{equation}
and the covariance matrix, $\Sigma$, for the final estimated vector is given by $\Sigma = \qty(L^T Q_{\alpha}^{-1} L)^{-1}$, such that $L = B^{-1} A B_2^{-1} G$.

\section{Numerical Results}

To evaluate the performance of the localization algorithm for initial orbit determination, we consider a simulated scenario with the initial target position and velocity given as

\begin{equation}
    \centering
    \begin{aligned}
    \mathbf{x} &= \begin{bmatrix}
        -2370406.31406129 \\
        -3691689.10408981 \\
        4901428.8809492
    \end{bmatrix} m \\
    \mathbf{v} &= \begin{bmatrix}
        -3931.046491 \\
        6498.676921 \\
        4665.980697
    \end{bmatrix} m/s.
    \end{aligned}
\end{equation}
We consider a multi-static radar system with three transmitters and five receivers, whose geographical coordinates are presented in Table~\ref{tab:lat-lon-table}, as well as the carrier frequencies of the transmitter antennas.
The covariance matrix of the measurements is given by $Q_{\alpha} = \sigma_t^2 \cdot \mbox{diag}\qty(\mathbf{I}_{MN}, 10^{11} \mathbf{I}_{MN})$. 

\renewcommand{\arraystretch}{1.5}
\begin{table}[h!]
    \centering
    \caption{Geographical coordinates of the transmitters and receivers of the considered multi-static radar system, as well as the carrier frequencies of the transmitter antennas.}
    \label{tab:lat-lon-table}
    \begin{tabular}{|c|c|c|c|}
    \hline
    & Latitude & Longitude & $f_c$ \\
    \hline
    $\mathbf{t}_1$ & 37.182º & -5.605º & $1215$ MHz \\
    \hline
    $\mathbf{t}_2$ & 44.335º & 7.638º & $1280$ MHz \\
    \hline
    $\mathbf{t}_3$ & 51.616º & 7.129º & $1330$ MHz \\
    \hline
    $\mathbf{s}_1$ & 40.000º & −3.600º & $-$ \\
    \hline
    $\mathbf{s}_2$ & 42.000º & 2.300º & $-$ \\
    \hline
    $\mathbf{s}_3$ & 46.000º & 4.300º & $-$ \\
    \hline
    $\mathbf{s}_4$ & 49.300º & -1.300º & $-$ \\
    \hline
    $\mathbf{s}_5$ & 42.000º & 6.300º & $-$ \\
    \hline
    \end{tabular}
\end{table}

We test the performance for different noise levels $\sigma_t$. We simulate noisy time delay and Doppler shift measurements by adding Gaussian noise with zero mean and standard deviation $\sigma_t$ and $\sqrt{10^{11}} \sigma_t$, respectively. We then use these noisy measurements to obtain range and range-rate for the trilateration method, under the assumption that they are independent of each other. For each level, we perform $S = 1000$ runs and we compute the Root Mean Squared Error (RMSE) for the position and velocity estimation as

\begin{equation}
    \centering
    \begin{aligned}
    \mbox{RMSE} = \sqrt{ \frac{1}{S} \sum_{i=1}^{S} \norm{\Hat{x} - x^0}^2 \ },
    \end{aligned}
\end{equation}
where $\Hat{x}$ denotes the estimated position or velocity, and $x^0$ denotes the true value. We compare the accuracy obtained in both stages of the algorithm, the estimation stage and the final result after the correction stage, with the Cramér-Rao Lower Bound (CRLB) accuracy and the trilateration method.

\begin{figure}[ht!]
    \centering
    \includegraphics[scale=0.55]{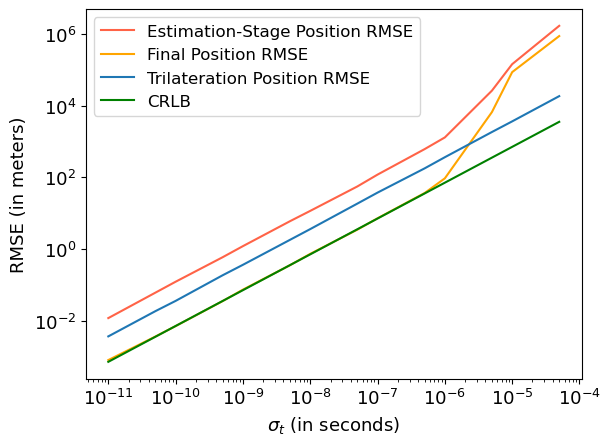}
    \caption{Root Mean Squared Error (RMSE) of position estimation. For small noise (up to $\sigma_t \approx 10^{-6} s$), the final estimate for the position of the object is able to attain CRLB accuracy. The estimation stage of the proposed algorithm is less accurate than the trilateration method. However, we are able to obtain a more accurate estimation of the velocity after the correction step.}
    \label{fig:position_MSE}
\end{figure}

\begin{figure}[ht!]
    \centering
    \includegraphics[scale=0.55]{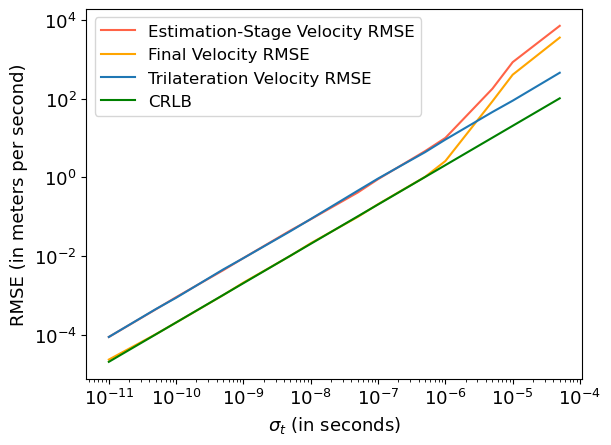}
    \caption{Root Mean Squared Error (RMSE) of velocity estimation. For small noise (up to $\sigma_t \approx 10^{-6} s$), the final estimate for the velocity of the object is able to attain CRLB accuracy. The estimation stage of the proposed algorithm is able to attain the same accuracy as the trilateration method. However, we are able to obtain a more accurate estimation of the velocity after the correction step.}
    \label{fig:velocity_MSE}
\end{figure}

In Figs.~\ref{fig:position_MSE} and~\ref{fig:velocity_MSE} it is possible to assess how the error behaves as $\sigma_t$ increases. It is clear that for small noise (up to $\sigma_t \approx 10^{-6} s$), the final estimate of our estimator, for both position and velocity, is able to attain CRLB accuracy. We see that for the position estimation, the trilateration method is more accurate than the result from our estimation stage, and for the velocity estimation, both achieve the same accuracy (up to $\sigma_t \approx 10^{-6} s$). However, after the correction stage of our approach, we see that the final result achieves better accuracy than the trilateration method.

In Table~\ref{tab:rmse-table}, we can analyze the Root Mean Squared Error of the position estimation by the trilateration method and our proposed method (WLS) for different levels of noise. For small noise (up to $\sigma_t \approx 10^{-6} s$), we see that the proposed method is able to attain an error which, generally, is one order of magnitude smaller than the error obtained by the trilateration approach.

For example, considering the same accuracy of clocks used in Global Positioning System (GPS) satellites around $10 ns$ (or $10^{-8} s$)~\cite{ostro1993planetary}, our approach is able to determine the position of the satellite with an error of $72 cm$, while the trilateration method presents an error of $3.59 m$.

\begin{table}
    \centering
    \caption{Root Mean Squared Error of the position estimation by the trilateration method and our proposed method (WLS) for different levels of noise.}
    \label{tab:rmse-table}
    \begin{tabular}{|c|c|c|}
    \hline
    $\sigma_t (s)$ & WLS & Trilaterion \\
    \hline
    $10^{-11}$ & $\mathbf{7.93 \times 10^{-4}} \, m$ & $3.63 \times 10^{-3} \, m$ \\
    \hline
    $10^{-10}$ & $\mathbf{7.04 \times 10^{-3}} \, m$ & $3.55 \times 10^{-2} \, m$ \\
    \hline
    $10^{-9}$ & $\mathbf{7.33 \times 10^{-2}} \, m$ & $3.61 \times 10^{-1} \, m$ \\
    \hline
    $10^{-8}$ & $\mathbf{7.31 \times 10^{-1}} \, m$ & $3.59 \, m$ \\
    \hline
    $10^{-7}$ & $\mathbf{7.18} \, m$ & $3.74 \times 10^{1} \, m$ \\
    \hline
    $10^{-6}$ & $\mathbf{9.37 \times 10^{1}} \, m$ & $3.64 \times 10^{2} \, m$ \\
    \hline
    \end{tabular}
\end{table}

\subsection{Impact of bias on the estimator}

By removing the quadratic noise terms, we introduce bias in the estimation, which we consider negligible for small noise. From the experiments, we can observe that the estimator is able to attain CRLB accuracy, i.e., the accuracy attained by the unbiased estimator with minimum variance~\cite{kay1993fundamentals}. We must now assess experimentally the impact of bias on the estimator, so that we can ascertain the validity of comparing our variance with the CRLB.

\begin{figure}[ht!]
    \centering
    \begin{subfigure}[b]{0.5\textwidth}
    \centering
    \includegraphics[scale=0.5]{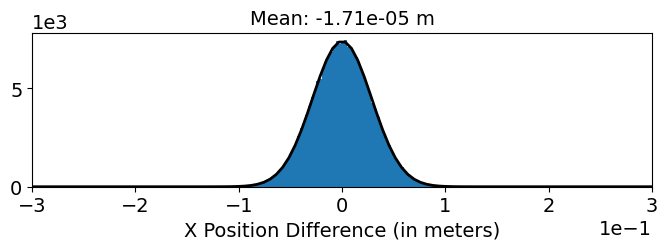}
    \caption{}
    \end{subfigure}
    \begin{subfigure}[b]{0.5\textwidth}
    \centering
    \includegraphics[scale=0.5]{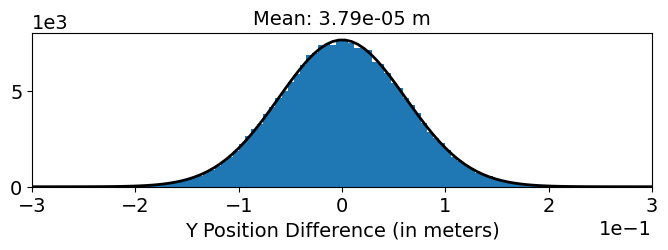}
    \caption{}
    \end{subfigure}
    \begin{subfigure}[b]{0.5\textwidth}
    \centering
    \includegraphics[scale=0.5]{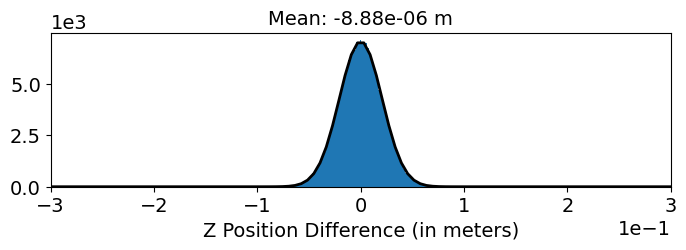}
    \caption{}
    \end{subfigure}
    \caption{Impact of bias in position estimation. For each axis, we can observe the histogram with the mean difference errors for the position estimation. We can see that for each component, the distribution seems to follow a Gaussian distribution with a mean close to zero, indicating a small impact of bias on the estimator.}
    \label{fig:position_mean_difference}
\end{figure}

\begin{figure}[ht!]
    \centering
    \begin{subfigure}[b]{0.5\textwidth}
    \centering
    \includegraphics[scale=0.5]{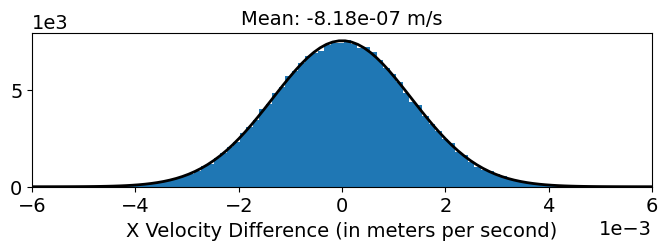}
    \caption{}
    \end{subfigure}
    \begin{subfigure}[b]{0.5\textwidth}
    \centering
    \includegraphics[scale=0.5]{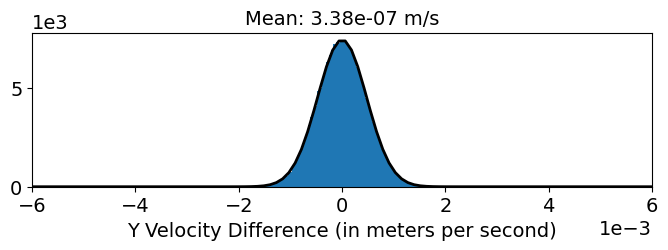}
    \caption{}
    \end{subfigure}
    \begin{subfigure}[b]{0.5\textwidth}
    \centering
    \includegraphics[scale=0.5]{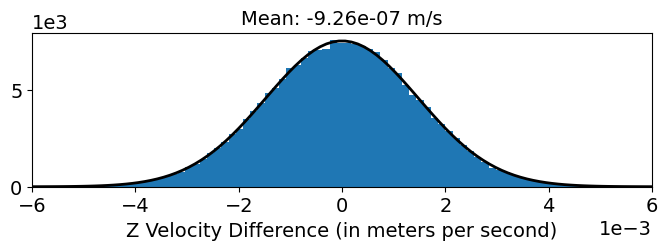}
    \caption{}
    \end{subfigure}
    \caption{Impact of bias in velocity  estimation. For each axis, we can observe the histogram with the mean difference errors for the velocity estimation. We can see that for each component, the distribution seems to follow a Gaussian distribution with a mean close to zero, indicating a small impact of bias on the estimator.}
    \label{fig:velocity_mean_difference}
\end{figure}

We performed $2 \times 10^{5}$ simulations for $\sigma_t = 10^{-9} s$ and computed, for each axis ($x, y, z$) of the position and velocity, the mean difference between our estimations and the true value, $\Hat{x}_i - {x_i}^0$, where $\Hat{x}_i$ denotes the $i$-th component of the position or velocity estimation and ${x_i}^0$ denotes the $i$-th component of the true value. In Figs.~\ref{fig:position_mean_difference} and~\ref{fig:velocity_mean_difference}, we can observe that the mean differences, for both position and velocity, seem to follow a Gaussian distribution with a mean close to zero, therefore strongly indicating a small impact of bias on the estimator.

In this way, we see that the bias is up to four orders of magnitude smaller than the error of the estimation and previously demonstrated that our estimator is able to approximate the same accuracy as the minimum variance unbiased estimator.

\subsection{Uncertainty Quantification}

We also compare the uncertainty quantification between our proposed method and the trilateration method, i.e., to compare how large the extracted covariance matrices are from each method. Given the critical role of satellites for society, it is important that the uncertainty associated with the estimation of the Resident Space Object (RSO) is taken into account and accurately modeled in order to assist the decision-making process of collision avoidance and trajectory tracking~\cite{klinkrad2005collision,klinkrad2006space,kleinig2022collision,bonnal2020just,reiland2021assessing,poore2016covariance}.

Similarly to the previous experiment, we performed $2 \times 10^5$ simulations for $\sigma_t = 10^{-9} s$ and computed, for each axis ($x, y, z$) of the position and velocity, the mean difference between each direction's standard deviation, $\sigma^{WLS}_{i} - \sigma^{Tri}_{i}$, for $i \in \{x, y, z\}$, obtained from the covariance matrix of each method, where $\sigma^{WLS}_i$ denotes the standard deviation of the $i$ direction from our proposed method and $\sigma^{Tri}_i$ denotes the standard deviation of the $i$ direction from trilateration method.

\begin{figure}[ht!]
    \centering
    \begin{subfigure}[b]{0.5\textwidth}
    \centering
    \includegraphics[scale=0.5]{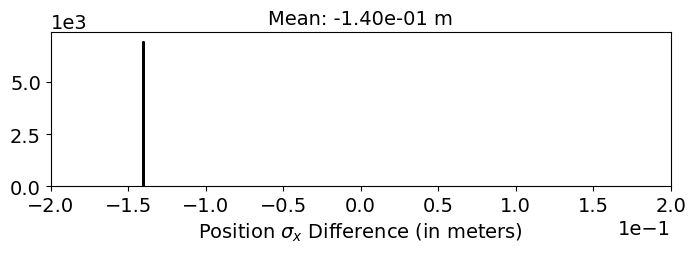}
    \caption{}
    \end{subfigure}
    \begin{subfigure}[b]{0.5\textwidth}
    \centering
    \includegraphics[scale=0.5]{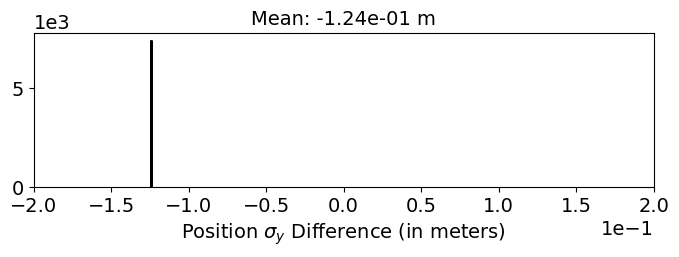}
    \caption{}
    \end{subfigure}
    \begin{subfigure}[b]{0.5\textwidth}
    \centering
    \includegraphics[scale=0.5]{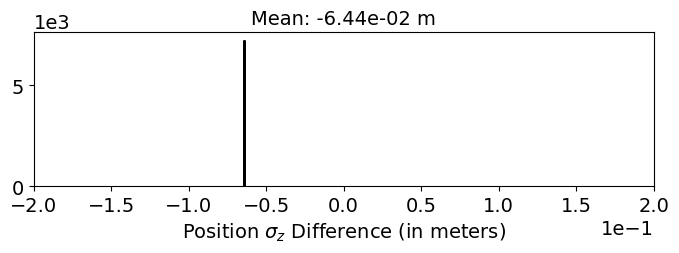}
    \caption{}
    \end{subfigure}
    \caption{Difference in position's uncertainty quantification. For each axis, we can observe the histogram with the mean difference between the standard deviations for the position estimation, of the trilateration approach and our proposed method. We can see that the uncertainty quantified from the proposed method is approximately one order of magnitude smaller than from the trilateration method.}
    \label{fig:position_sigma_difference}
\end{figure}
\begin{figure}[ht!]
    \centering
    \begin{subfigure}[b]{0.5\textwidth}
    \centering
    \includegraphics[scale=0.5]{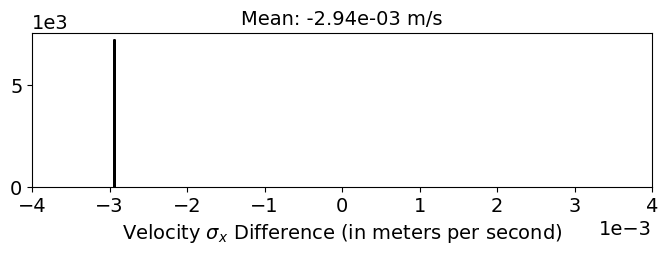}
    \caption{}
    \end{subfigure}
    \begin{subfigure}[b]{0.5\textwidth}
    \centering
    \includegraphics[scale=0.5]{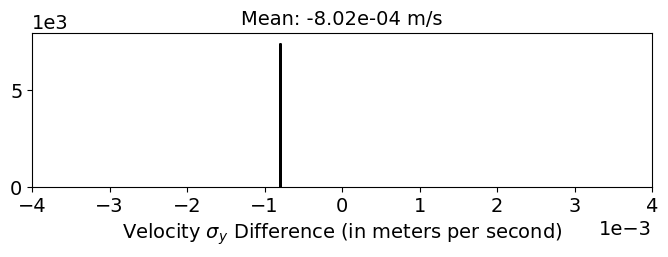}
    \caption{}
    \end{subfigure}
    \begin{subfigure}[b]{0.5\textwidth}
    \centering
    \includegraphics[scale=0.5]{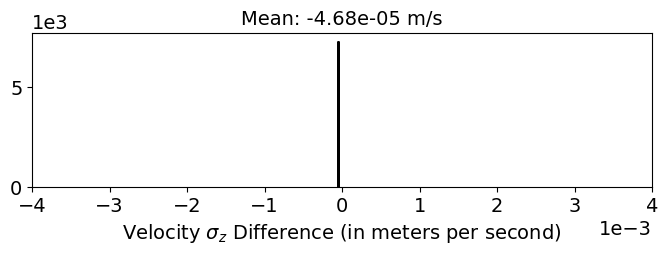}
    \caption{}
    \end{subfigure}
    \caption{Difference in velocity's uncertainty quantification. For each axis, we can observe the histogram with the mean difference between the standard deviations for the velocity. The difference is less or equal to two orders of magnitude for the velocity estimation. Yet, our proposed method quantifies a smaller uncertainty in every direction than the trilateration method.}
    \label{fig:velocity_sigma_difference}
\end{figure}
In Figs.~\ref{fig:position_sigma_difference} and~\ref{fig:velocity_sigma_difference}, we can observe that the mean differences between the standard deviations, for both position and velocity, are always less than zero. This indicates that our proposed method returns a lower uncertainty on the estimation than the trilateration method for both position and velocity. 

For the position estimation, our method is able to model the uncertainty one order of magnitude smaller for every direction. For the velocity estimation, the difference is not as large, however, it still presents lower uncertainty than the trilateration method.

\begin{figure}[ht]
    \centering
    \includegraphics[scale=0.8]{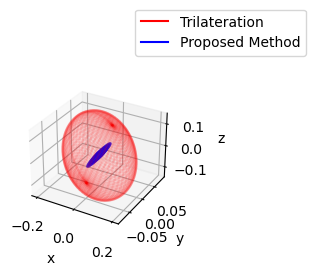}
    \caption{Uncertainty ellipsoid. Comparison of uncertainty ellipsoid between the proposed method and trilateration method for the position estimation.}
    \label{fig:confidence_ellipsoid}
\end{figure}
In Fig.~\ref{fig:confidence_ellipsoid}, we can observe a significant difference in the ellipsoid volume between our method and the trilateration. This qualitatively demonstrates that our approach returns a ``tighter'' estimation for the position.

\section{Conclusions and Future Work}

With this work, we present a non-iterative estimator for the problem of initial orbit determination for Low-Earth Orbit, providing the complete state vector of the object (position and velocity) and the corresponding associated covariance matrix. We demonstrate that, for small Gaussian measurement noise, the estimator is able to attain CRLB accuracy. Our method does not need to propagate a reference trajectory, bypassing simplifications of the nonlinear dynamical system and avoiding time-consuming steps such as ODE integration.

We experimentally assess the bias introduced on our estimator for disregarding the quadratic noise terms and observed that it has a very small impact on the estimation, achieving the same accuracy as the minimum variance unbiased estimator. We also compared the associated error between our method and trilateration. Both position and velocity error from our method is smaller than the trilateration method, presenting a more ``tighter'' estimation and positively complementing the decision-making process of collision avoidance and trajectory tracking because it allows a better awareness of the object's state.

As demonstrated by Yang et al.~\cite{yang2016moving}, through CRLB analysis, the accuracy of the estimate for the object's position can be increased by Doppler measurements when the object is closer to the transmitter or receivers. Future work could evaluate the performance of this estimator for other types of orbit, such as Medium-Earth Orbit (MEO) and Geostationary Orbit (GEO), where the distance from the transmitters and receivers can have a significant impact on the accuracy of the estimator.

Some work has been developed for satellite-to-satellite tracking~\cite{liu2001orbit}, with studies of the applicability and accuracy of the trilateration method~\cite{vonbun1978orbit}. In this work, we demonstrate that our approach is more accurate than the trilateration method, therefore it would be interesting to study and compare our approach for the problem of satellite-to-satellite tracking. Also, our method is more sensitive to ground stations that are very close to each other as the angle between the transmitter-target and receiver-target directions becomes smaller. For satellite-to-satellite tracking, with the combined use of ground stations and geostationary satellites, this problem is immediately overcome.

%
% ---- Appendices ----

%\begin{appendices}
\appendices

\section{Estimation Stage - Noise Terms Relation}
\label{app:stage1-noise-terms-relation}

In this section, we present a detailed derivation of the expressions that relate the differential delay and Doppler shift measurements with the position, $\mathbf{x}$, and velocity, $\mathbf{v}$, variables. We start by deriving the expression presented in~\eqref{eq:differential-delay-noise-relaxation}. Starting by rearranging the equation~\eqref{eq:differential-delay-model} as

\begin{equation}
    \label{eq:norm-sj-transformation}
    \centering
    \begin{aligned}
    c \tau_{ij}^0 - \norm{\mathbf{x} - \mathbf{t}_i} = \norm{\mathbf{x} - \mathbf{s}_j}
    \end{aligned}
\end{equation}
and substituting $\tau_{ij}^0 = \tau_{ij} - \Delta \tau_{ij}$, we have

\begin{equation}
    \centering
    \begin{aligned}
    & c \tau_{ij} - \norm{\mathbf{x} - \mathbf{t}_i} = \norm{\mathbf{x} - \mathbf{s}_j} + c \Delta \tau_{ij}
    \end{aligned}
\end{equation}
By squaring both sides, we obtain

\begin{equation}
    \centering
    \begin{aligned}
    & \qty( c \tau_{ij} - \norm{\mathbf{x} - \mathbf{t}_i} )^2 = \qty( \norm{\mathbf{x} - \mathbf{s}_j} + c \Delta \tau_{ij} )^2 \iff \\
    \\
    & c^2 \tau_{ij}^2 - 2 c \tau_{ij} \norm{\mathbf{x} - \mathbf{t}_i} + \norm{\mathbf{x}}^2 - 2 \, \mathbf{t}_i^T \mathbf{x} + \norm{\mathbf{t}_i}^2 \\
    &= \norm{\mathbf{x}}^2 - 2 \, {\mathbf{s}}_j^T {\mathbf{x}} + \norm{\mathbf{s}_j}^2 + 2 c \norm{\mathbf{x} - \mathbf{s}_j} \Delta \tau_{ij} \\
    &+ c^2 \Delta \tau_{ij}^2 \iff \\
    \\
    & c^2 \tau_{ij}^2 + \norm{\mathbf{t}_i}^2 - \norm{\mathbf{s}_j}^2 - 2 \qty( \mathbf{t}_i - \mathbf{s}_j )^T \mathbf{x} \\
    &- 2 c \tau_{ij} \norm{\mathbf{x} - \mathbf{t}_i} = 2 c \norm{\mathbf{x} - \mathbf{s}_j} \Delta \tau_{ij} + c^2 \Delta \tau_{ij}^2
    \end{aligned}
\end{equation}
By ignoring the second-order noise term on the right side of the equation, we arrive at the final expression

\begin{equation}
    \centering
    \begin{aligned}
    &c^2 \tau_{ij}^2 + \norm{\mathbf{t}_i}^2 - \norm{\mathbf{s}_j}^2 - 2 \qty( \mathbf{t}_i - \mathbf{s}_j )^T \mathbf{x} \\
    &- 2 c \tau_{ij} \norm{\mathbf{x} - \mathbf{t}_i} \approx 2 c \norm{\mathbf{x} - \mathbf{s}_j} \Delta \tau_{ij}
    \end{aligned}
\end{equation}
To derive the expression presented in~\eqref{eq:doppler-shift-noise-relation}, we follow the same approach but for the doppler shift model defined as 

\begin{equation}
    \centering
    \begin{aligned}
    f_{ij}^0 = \frac{f_{c,i}}{c} \qty( \rho_{\mathbf{x}, \mathbf{t}_i}^T \mathbf{v} + \rho_{\mathbf{x}, \mathbf{s}_j}^T \mathbf{v} )
    \end{aligned}
\end{equation}
By substituting $f_{ij}^0 = f_{ij} - \Delta f_{ij}$ and multiplying both sides by $c \norm{\mathbf{x} - \mathbf{t}_i} \norm{\mathbf{x} - \mathbf{s}_j}$, we have

\begin{equation}
    \centering
    \begin{aligned}
    &\norm{\mathbf{x} - \mathbf{t}_i} \norm{\mathbf{x} - \mathbf{s}_j} c f_{ij} - \norm{\mathbf{x} - \mathbf{t}_i} \norm{\mathbf{x} - \mathbf{s}_j} c \Delta f_{ij} = \\
    &= f_{c,i} \qty( \norm{\mathbf{x} - \mathbf{s}_j} \qty(\mathbf{x} - \mathbf{t}_i)^T \mathbf{v} + \norm{\mathbf{x} - \mathbf{t}_i}  \qty(\mathbf{x} - \mathbf{s}_j)^T \mathbf{v} )
    \end{aligned}
\end{equation}
We can substitute $\norm{\mathbf{x} - \mathbf{s}_j}$ through~\eqref{eq:norm-sj-transformation}. By substituting all the terms $\norm{\mathbf{x} - \mathbf{s}_j}$ except the one associated with the error term, we obtain

\begin{equation}
    \centering
    \begin{aligned}
    &\norm{\mathbf{x} - \mathbf{t}_i} \qty(c \tau_{ij}^0 - \norm{\mathbf{x} - \mathbf{t}_i}) c f_{ij} - \norm{\mathbf{x} - \mathbf{t}_i} \norm{\mathbf{x} - \mathbf{s}_j} c \Delta f_{ij} = \\
    &= f_{c,i} \left( \qty(c \tau_{ij}^0 - \norm{\mathbf{x} - \mathbf{t}_i}) \qty(\mathbf{x} - \mathbf{t}_i)^T \mathbf{v} + \norm{\mathbf{x} - \mathbf{t}_i}  \qty(\mathbf{x} - \mathbf{s}_j)^T \mathbf{v} \right).
    \end{aligned}
\end{equation}
By dividing both sides of the equation by $\norm{\mathbf{x} - \mathbf{t}_i}$, we arrive at

\begin{equation}
    \centering
    \begin{aligned}
    &\qty(c \tau_{ij}^0 - \norm{\mathbf{x} - \mathbf{t}_i}) c f_{ij} - \norm{\mathbf{x} - \mathbf{s}_j} c \Delta f_{ij} = \\
    &= f_{c,i} \qty( \qty(c \tau_{ij}^0 - \norm{\mathbf{x} - \mathbf{t}_i}) \rho_{\mathbf{x}, \mathbf{t}_i}^T \mathbf{v} + \qty(\mathbf{x} - \mathbf{s}_j)^T \mathbf{v} ).
    \end{aligned}
\end{equation}
Similarly to before, if we substitute $\tau_{ij}^0 = \tau_{ij} - \Delta \tau_{ij}$, we obtain

\begin{equation}
    \centering
    \begin{aligned}
    &\qty(c \tau_{ij} - c \Delta \tau_{ij} - \norm{\mathbf{x} - \mathbf{t}_i}) c f_{ij} - \norm{\mathbf{x} - \mathbf{s}_j} c \Delta f_{ij} = \\
    &= f_{c,i} \left( \qty(c \tau_{ij} - c \Delta \tau_{ij} - \norm{\mathbf{x} - \mathbf{t}_i}) \rho_{\mathbf{x}, \mathbf{t}_i}^T \mathbf{v} \right. \\
    &+ \left. \qty(\mathbf{x} - \mathbf{s}_j)^T \mathbf{v} \right) \iff \\
    \\
    & c^2 f_{ij} \tau_{ij} - c^2 f_{ij} \Delta \tau_{ij} - c f_{ij} \norm{\mathbf{x} - \mathbf{t}_i} - \norm{\mathbf{x} - \mathbf{s}_j} c \Delta f_{ij} \\ 
    &= f_{c,i} c \tau_{ij} \rho_{\mathbf{x}, \mathbf{t}_i}^T {\mathbf{v}} - f_{c,i} c \rho_{\mathbf{x}, \mathbf{t}_i}^T {\mathbf{v}} \Delta \tau_{ij} + \\
    &f_{c,i} \qty(-\mathbf{x} + \mathbf{t}_i + \mathbf{x} - \mathbf{s}_j)^T \mathbf{v} \iff \\
    \\
    & c^2 f_{ij} \tau_{ij} - f_{c,i} \qty(\mathbf{t}_i - \mathbf{s}_j)^T {\mathbf{v}} - c f_{ij} \norm{\mathbf{x} - \mathbf{t}_i} \\
    &- f_{c,i} c \tau_{ij} \rho_{\mathbf{x}, \mathbf{t}_i}^T {\mathbf{v}} = c^2 f_{ij} \Delta \tau_{ij} - f_{c,i} c \rho_{\mathbf{x}, \mathbf{t}_i}^T {\mathbf{v}} \Delta \tau_{ij} \\
    &+ \norm{\mathbf{x} - \mathbf{s}_j} c \Delta f_{ij} \iff \\
    \\
    & c^2 f_{ij} \tau_{ij} - f_{c,i} \qty(\mathbf{t}_i - \mathbf{s}_j)^T {\mathbf{v}} - c f_{ij} \norm{\mathbf{x} - \mathbf{t}_i} \\
    &- f_{c,i} c \tau_{ij} \rho_{\mathbf{x}, \mathbf{t}_i}^T {\mathbf{v}} = c^2 \Delta \tau_{ij} \qty( f_{ij} - \frac{f_{c,i}}{c} \rho_{\mathbf{x}, \mathbf{t}_i}^T {\mathbf{v}} ) \\
    &+ \norm{\mathbf{x} - \mathbf{s}_j} c \Delta f_{ij}
    \end{aligned}
\end{equation}
By manipulating~\eqref{eq:doppler-shift-model} and~\eqref{eq:doppler-shift-measurement}, we arrive at $f_{ij} - \frac{f_{c,i}}{c} \rho_{\mathbf{x}, \mathbf{t}_i}^T {\mathbf{v}} = \Delta f_{ij} + \frac{f_{c,i}}{c} \rho_{\mathbf{x}, \mathbf{s}_j}^T \mathbf{v}$. By using this result to substitute in the right side of the equation, we have

\begin{equation}
    \centering
    \begin{aligned}
    & c^2 f_{ij} \tau_{ij} - f_{c,i} \qty(\mathbf{t}_i - \mathbf{s}_j)^T {\mathbf{v}} - c f_{ij} \norm{\mathbf{x} - \mathbf{t}_i} - f_{c,i} c \tau_{ij} \rho_{\mathbf{x}, \mathbf{t}_i}^T \mathbf{v} = \\
    &= c^2 \Delta \tau_{ij} \qty( \Delta f_{ij} + \frac{f_{c,i}}{c} \rho_{\mathbf{x}, \mathbf{s}_j}^T \mathbf{v} ) + \norm{\mathbf{x} - \mathbf{s}_j} c \Delta f_{ij} \iff \\
    \\
    & c^2 f_{ij} \tau_{ij} - f_{c,i} \qty(\mathbf{t}_i - \mathbf{s}_j)^T {\mathbf{v}} - c f_{ij} \norm{\mathbf{x} - \mathbf{t}_i} - f_{c,i} c \tau_{ij} \rho_{\mathbf{x}, \mathbf{t}_i}^T \mathbf{v} = \\
    &= c^2 \Delta \tau_{ij} \Delta f_{ij} + c f_{c,i} \rho_{\mathbf{x}, \mathbf{s}_j}^T {\mathbf{v}} \Delta \tau_{ij} + \norm{\mathbf{x} - \mathbf{s}_j} c \Delta f_{ij}
    \end{aligned}
\end{equation}
By ignoring the second-order noise term on the right side of the equation and multiplying both sides by 2, we arrive at the final expression

\begin{equation}
    \centering
    \begin{aligned}
    & 2 c^2 f_{ij} \tau_{ij} - 2 f_{c,i} \qty(\mathbf{t}_i - \mathbf{s}_j)^T {\mathbf{v}} - 2 c f_{ij} \norm{\mathbf{x} - \mathbf{t}_i} \\
    &- 2 c f_{c,i} \tau_{ij} \rho_{\mathbf{x}, \mathbf{t}_i}^T {\mathbf{v}} = 2 c f_{c,i} \rho_{\mathbf{x}, \mathbf{s}_j}^T {\mathbf{v}} \Delta \tau_{ij} \\
    &+ 2 c \norm{\mathbf{x} - \mathbf{s}_j} \Delta f_{ij}
    \end{aligned}
\end{equation}

\section{Correction Stage - Noise Terms Relation}
\label{app:stage2-noise-terms-relation}

In this section, we present a detailed derivation of the expressions that relate the intermediate variables $\gamma_i$ and $\beta_i$ with the position, $\mathbf{x}$, and velocity, $\mathbf{v}$, variables. We start by deriving the expression presented in~\eqref{eq:stage2-gamma-intermediate-noise-terms-relation}. Remembering the intermediate variable $\gamma_i = \norm{\mathbf{x} - \mathbf{t}_i}$, we can define the estimated value as

\begin{equation}
    \centering
    \begin{aligned}
    &\Tilde{\gamma}_i = \gamma_i + \Delta \Tilde{\gamma}_i \iff \\
    &\Tilde{\gamma}_i - \Delta \Tilde{\gamma}_i = \norm{\mathbf{x} - \mathbf{t}_i}
    \end{aligned}
\end{equation}
where $\Delta \Tilde{\gamma}_i$ denotes the random error associated with the intermediate variable $\gamma_i$. By squaring both sides, we obtain

\begin{equation}
    \centering
    \begin{aligned}
    &\qty(\Tilde{\gamma}_i - \Delta \Tilde{\gamma}_i)^2 = \norm{\mathbf{x} - \mathbf{t}_i}^2 \iff \\
    &- 2 \Tilde{\gamma}_i \Delta \Tilde{\gamma}_i + \Delta \Tilde{\gamma}_i^2 = -\Tilde{\gamma}_i^2 +
    {\mathbf{x}}^T {\mathbf{x}} - 2 {\mathbf{t}}_i^T {\mathbf{x}} + {\mathbf{t}}_i^T {\mathbf{t}}_i
    \end{aligned}
\end{equation}
We can also define the estimated value for the position as having an associated error as $\Tilde{\mathbf{x}} = {\mathbf{x}} + \Delta {\mathbf{x}} \iff {\mathbf{x}} = \Tilde{\mathbf{x}} - \Delta {\mathbf{x}}$. By substituting the position variable, we have

\begin{equation}
    \centering
    \begin{aligned}
    &- 2 \Tilde{\gamma}_i \Delta \Tilde{\gamma}_i + \Delta \Tilde{\gamma}_i^2 = -\Tilde{\gamma}_i^2 +
    \qty(\Tilde{\mathbf{x}} - \Delta \mathbf{x})^T \qty(\Tilde{\mathbf{x}} - \Delta \mathbf{x}) \\
    &- 2 {\mathbf{t}}_i^T \qty(\Tilde{\mathbf{x}} - \Delta {\mathbf{x}}) + {\mathbf{t}}_i^T {\mathbf{t}}_i \iff \\
    & 2 \Tilde{\gamma}_i \Delta \Tilde{\gamma}_i - \Delta \Tilde{\gamma}_i^2 = \Tilde{\gamma}_i^2 - \Tilde{\mathbf{x}}^T \Tilde{\mathbf{x}} + 2 {\mathbf{t}}_i^T \Tilde{\mathbf{x}} - {\mathbf{t}}_i^T {\mathbf{t}}_i \\
    &+ 2 \qty(\Tilde{\mathbf{x}} - {\mathbf{t}}_i)^T \Delta {\mathbf{x}} - \Delta {\mathbf{x}}^2
    \end{aligned}
\end{equation}
By ignoring the second-order noise terms on both sides of the equation, we obtain the final expression

\begin{equation}
    \centering
    \begin{aligned}
    2 \Tilde{\gamma}_i \Delta \Tilde{\gamma}_i \approx \Tilde{\gamma}_i^2 - \Tilde{\mathbf{x}}^T \Tilde{\mathbf{x}} + 2 {\mathbf{t}}_i^T \Tilde{\mathbf{x}} - {\mathbf{t}}_i^T {\mathbf{t}}_i + 2 \qty(\Tilde{\mathbf{x}} - {\mathbf{t}}_i)^T \Delta \mathbf{x}
    \end{aligned}
\end{equation}
presented in~\eqref{eq:stage2-gamma-intermediate-noise-terms-relation}. To derive the expression presented in~\eqref{eq:stage2-beta-intermediate-noise-terms-relation}, we follow the same approach but for the intermediate variable $\beta_i = \rho_{\mathbf{x}, \mathbf{t}_i}^T \mathbf{v}$. We can define the estimated value as

\begin{equation}
    \centering
    \begin{aligned}
    &\Tilde{\beta}_i = \beta_i + \Delta \Tilde{\beta}_i \iff \\
    &\Tilde{\beta}_i - \Delta \Tilde{\beta}_i =  \frac{\qty({\mathbf{x}} - {\mathbf{t}}_i)^T}{\gamma_i} {\mathbf{v}} \iff \\
    &\qty(\Tilde{\beta}_i - \Delta \Tilde{\beta}_i) \gamma_i = {\mathbf{x}}^T {\mathbf{v}} - {\mathbf{t}}_i^T {\mathbf{v}}
    \end{aligned}
\end{equation}
Similarly, we can also define the estimated value for the variables $\gamma_i$, $\mathbf{x}$ and $\mathbf{v}$ as having an associated error. By substituting in the equation, we have

\begin{equation}
    \centering
    \begin{aligned}
    &\qty(\Tilde{\beta}_i - \Delta \Tilde{\beta}_i) \qty(\Tilde{\gamma}_i - \Delta \Tilde{\gamma}_i) = \qty(\Tilde{\mathbf{x}} - \Delta \mathbf{x})^T \qty(\Tilde{\mathbf{v}} - \Delta \mathbf{v}) \\
    &- {\mathbf{t}}_i^T \qty(\Tilde{\mathbf{v}} - \Delta {\mathbf{v}}) \iff \\
    &\Tilde{\beta}_i \Delta \Tilde{\gamma}_i + \Tilde{\gamma}_i \Delta \Tilde{\beta}_i - \Delta \Tilde{\beta}_i \Delta \Tilde{\gamma}_i = \Tilde{\beta}_i \Tilde{\gamma}_i - \Tilde{\mathbf{x}}^T \Tilde{\mathbf{v}} \\
    &+ {\mathbf{t}}_i^T \Tilde{\mathbf{v}} + \Tilde{\mathbf{v}}^T \Delta {\mathbf{x}} + \qty(\Tilde{\mathbf{x}} - {\mathbf{t}}_i)^T \Delta {\mathbf{v}} - \Delta {\mathbf{x}}^T \Delta {\mathbf{v}}
    \end{aligned}
\end{equation}
By ignoring the second-order noise terms on both sides of the equation, we obtain the final expression

\begin{equation}
    \centering
    \begin{aligned}
    \Tilde{\beta}_i \Delta \Tilde{\gamma}_i &+ \Tilde{\gamma}_i \Delta \Tilde{\beta}_i \approx \Tilde{\beta}_i \Tilde{\gamma}_i - \Tilde{\mathbf{x}}^T \Tilde{\mathbf{v}} + {\mathbf{t}}_i^T \Tilde{\mathbf{v}} \\
    &+ \Tilde{\mathbf{v}}^T \Delta {\mathbf{x}} + \qty(\Tilde{\mathbf{x}} - {\mathbf{t}}_i)^T \Delta {\mathbf{v}}
    \end{aligned}
\end{equation}
presented in~\eqref{eq:stage2-beta-intermediate-noise-terms-relation}.

%
% ---- Acknowledgments ----

\acknowledgments
The authors would like to express their gratitude to Engineer M\'{a}rcio Menezes, from Neuraspace.

This work was partially supported by NOVA LINCS (UIDB/04516/2020)
with the financial support of FCT I.P. and Project “Artificial Intelligence Fights Space Debris” No C626449889-0046305 co-funded by Recovery and Resilience Plan and NextGeneration EU Funds, www.recuperarportugal.gov.pt.\\
\begin{center}
    \includegraphics[scale=0.1]{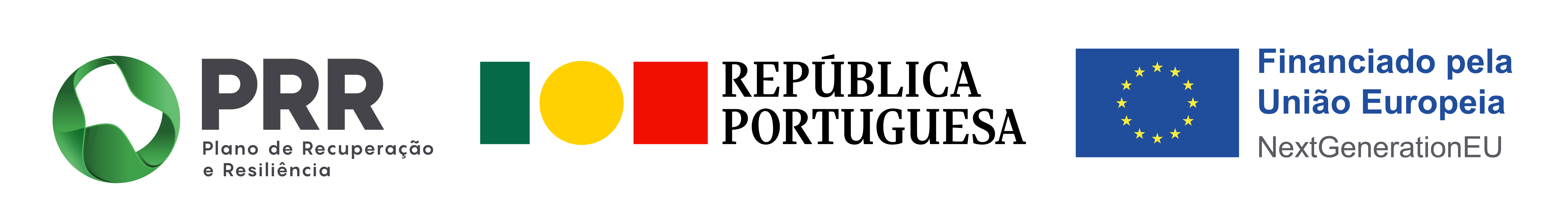}
\end{center}


\bibliographystyle{IEEEtran}
\bibliography{bibliografia}

\begin{thebibliography}{10}
\providecommand{\url}[1]{#1}
\csname url@samestyle\endcsname
\providecommand{\newblock}{\relax}
\providecommand{\bibinfo}[2]{#2}
\providecommand{\BIBentrySTDinterwordspacing}{\spaceskip=0pt\relax}
\providecommand{\BIBentryALTinterwordstretchfactor}{4}
\providecommand{\BIBentryALTinterwordspacing}{\spaceskip=\fontdimen2\font plus
\BIBentryALTinterwordstretchfactor\fontdimen3\font minus
  \fontdimen4\font\relax}
\providecommand{\BIBforeignlanguage}[2]{{%
\expandafter\ifx\csname l@#1\endcsname\relax
\typeout{** WARNING: IEEEtran.bst: No hyphenation pattern has been}%
\typeout{** loaded for the language `#1'. Using the pattern for}%
\typeout{** the default language instead.}%
\else
\language=\csname l@#1\endcsname
\fi
#2}}
\providecommand{\BIBdecl}{\relax}
\BIBdecl

\bibitem{esa-numbers}
\BIBentryALTinterwordspacing
{ESA}'s Space Debris~Office. (2023) Space debris by the numbers. [Online].
  Available:
  \url{https://www.esa.int/Space_Safety/Space_Debris/Space_debris_by_the_numbers}
\BIBentrySTDinterwordspacing

\bibitem{schutz2004statistical}
B.~Schutz, B.~Tapley, and G.~H. Born, \emph{Statistical orbit
  determination}.\hskip 1em plus 0.5em minus 0.4em\relax Elsevier, 2004.

\bibitem{vallado2001fundamentals}
D.~A. Vallado, \emph{Fundamentals of astrodynamics and applications}.\hskip 1em
  plus 0.5em minus 0.4em\relax Springer Science \& Business Media, 2001,
  vol.~12.

\bibitem{lam2010analysis}
Q.~Lam, D.~Junker, D.~Anhalt, and D.~Vallado, ``Analysis of an extended
  {K}alman filter based orbit determination system,'' in \emph{AIAA Guidance,
  Navigation, and Control Conference}, 2010, p. 7600.

\bibitem{pardal2011robustness}
P.~Pardal, H.~Kuga, and R.~V. de~Moraes, ``Robustness assessment between sigma
  point and extended kalman filter for orbit determination,'' \emph{Journal of
  {A}erospace {E}ngineering, {S}ciences and {A}pplications}, vol.~3, no.~3, pp.
  35--44, 2011.

\bibitem{krener2003convergence}
A.~J. Krener, \emph{The convergence of the extended {K}alman filter}.\hskip 1em
  plus 0.5em minus 0.4em\relax Springer, 2003.

\bibitem{poore2016covariance}
A.~B. Poore, J.~M. Aristoff, J.~T. Horwood, R.~Armellin, W.~T. Cerven,
  Y.~Cheng, C.~M. Cox, R.~S. Erwin, J.~H. Frisbee, M.~D. Hejduk \emph{et~al.},
  ``Covariance and uncertainty realism in space surveillance and tracking,''
  Numerica Corporation Fort Collins United States, Tech. Rep., 2016.

\bibitem{julier1997new}
S.~J. Julier and J.~K. Uhlmann, ``New extension of the {K}alman filter to
  nonlinear systems,'' in \emph{Signal processing, sensor fusion, and target
  recognition VI}, vol. 3068.\hskip 1em plus 0.5em minus 0.4em\relax Spie,
  1997, pp. 182--193.

\bibitem{einicke2012smoothing}
G.~Einicke, \emph{Smoothing, filtering and prediction: {E}stimating the past,
  present and future}.\hskip 1em plus 0.5em minus 0.4em\relax BoD--Books on
  Demand, 2012.

\bibitem{escobal1970methods}
P.~Escobal, ``Methods of orbit determination,'' \emph{Methods of orbit
  determination}, 1970.

\bibitem{gibbs1889determination}
J.~W. Gibbs, \emph{On the determination of elliptic orbits from three complete
  observations}.\hskip 1em plus 0.5em minus 0.4em\relax National Academy of
  Sciences, 1889, vol.~4.

\bibitem{herrick1971astrodynamics}
S.~Herrick, ``Astrodynamics: Orbit determination, space navigation,''
  \emph{Celestial Mechanics}, vol.~1, pp. 90--93, 1971.

\bibitem{kaushik2016statistical}
A.~S. Kaushik, ``A statistical comparison between gibbs and herrick-gibbs orbit
  determination methods,'' Ph.D. dissertation, 2016.

\bibitem{gooding1996new}
R.~H. Gooding, ``A new procedure for the solution of the classical problem of
  minimal orbit determination from three lines of sight,'' \emph{Celestial
  Mechanics and Dynamical Astronomy}, vol.~66, pp. 387--423, 1996.

\bibitem{izzo2015revisiting}
D.~Izzo, ``Revisiting lambert’s problem,'' \emph{Celestial Mechanics and
  Dynamical Astronomy}, vol. 121, pp. 1--15, 2015.

\bibitem{gooding1988solution}
R.~Gooding, \emph{On the solution of Lambert's orbital boundary-value
  problem}.\hskip 1em plus 0.5em minus 0.4em\relax Royal Aerospace
  Establishment, 1988.

\bibitem{gooding1990procedure}
R.~H. Gooding, ``A procedure for the solution of lambert's orbital
  boundary-value problem,'' \emph{Celestial Mechanics and Dynamical Astronomy},
  vol.~48, no.~2, pp. 145--165, 1990.

\bibitem{lancaster1969unified}
E.~Lancaster, \emph{A unified form of Lambert's theorem}.\hskip 1em plus 0.5em
  minus 0.4em\relax National Aeronautics and Space Administration, 1969.

\bibitem{shang2019VSA}
H.~Shang, D.~Chen, H.~Cao, T.~Fu, and M.~Gao, ``Initial orbit determination
  using very short arc data based on double-station observation,'' \emph{IEEE
  Transactions on Aerospace and Electronic Systems}, vol.~55, no.~4, pp.
  1596--1611, 2019.

\bibitem{qu2022VSA}
J.~Qu, D.~Chen, H.~Cao, T.~Fu, and S.~Zhang, ``Initial orbit determination
  method for low earth orbit objects using too-short arc based on bistatic
  radar,'' \emph{IEEE Access}, vol.~10, pp. 76\,766--76\,779, 2022.

\bibitem{hough2012precise}
M.~E. Hough, ``Precise orbit determination using satellite radar ranging,''
  \emph{Journal of Guidance, Control, and Dynamics}, vol.~35, no.~4, pp.
  1048--1058, 2012.

\bibitem{rui2015efficient}
L.~Rui and K.~Ho, ``Efficient closed-form estimators for multistatic sonar
  localization,'' \emph{IEEE Transactions on Aerospace and Electronic Systems},
  vol.~51, no.~1, pp. 600--614, 2015.

\bibitem{yang2016moving}
L.~Yang, L.~Yang, and K.~Ho, ``Moving target localization in multistatic sonar
  by differential delays and {D}oppler shifts,'' \emph{IEEE Signal Processing
  Letters}, vol.~23, no.~9, pp. 1160--1164, 2016.

\bibitem{einemo2015weighted}
M.~Einemo and H.~C. So, ``Weighted least squares algorithm for target
  localization in distributed {MIMO} radar,'' \emph{Signal Processing}, vol.
  115, pp. 144--150, 2015.

\bibitem{bishop2006pattern}
C.~M. Bishop and N.~M. Nasrabadi, \emph{Pattern {R}ecognition and {M}achine
  {L}earning}.\hskip 1em plus 0.5em minus 0.4em\relax Springer, 2006, vol.~4,
  no.~4.

\bibitem{kay1993fundamentals}
S.~M. Kay, \emph{Fundamentals of {S}tatistical {S}ignal {P}rocessing:
  {E}stimation {T}heory}.\hskip 1em plus 0.5em minus 0.4em\relax Prentice-Hall,
  Inc., 1993.

\bibitem{ostro1993planetary}
S.~J. Ostro, ``Planetary radar astronomy,'' \emph{Reviews of Modern Physics},
  vol.~65, no.~4, p. 1235, 1993.

\bibitem{klinkrad2005collision}
H.~Klinkrad, J.~Alarcon, and N.~Sanchez, ``Collision avoidance for operational
  esa satellites,'' in \emph{4th European Conference on Space Debris}, vol.
  587, 2005, p. 509.

\bibitem{klinkrad2006space}
H.~Klinkrad, \emph{Space debris: models and risk analysis}.\hskip 1em plus
  0.5em minus 0.4em\relax Springer Science \& Business Media, 2006.

\bibitem{kleinig2022collision}
T.~Kleinig, B.~Smith, and C.~Capon, ``Collision avoidance of satellites using
  ionospheric drag,'' \emph{Acta Astronautica}, 2022.

\bibitem{bonnal2020just}
C.~Bonnal, D.~McKnight, C.~Phipps, C.~Dupont, S.~Missonnier, L.~Lequette,
  M.~Merle, and S.~Rommelaere, ``Just in time collision avoidance--a review,''
  \emph{Acta Astronautica}, vol. 170, pp. 637--651, 2020.

\bibitem{reiland2021assessing}
N.~Reiland, A.~J. Rosengren, R.~Malhotra, and C.~Bombardelli, ``Assessing and
  minimizing collisions in satellite mega-constellations,'' \emph{Advances in
  Space Research}, vol.~67, no.~11, pp. 3755--3774, 2021.

\bibitem{liu2001orbit}
Y.-C. Liu and L.~Liu, ``Orbit determination using satellite-to-satellite
  tracking data,'' \emph{Chinese Journal of Astronomy and Astrophysics},
  vol.~1, no.~3, p. 281, 2001.

\bibitem{vonbun1978orbit}
F.~Vonbun, P.~Argentiero, and P.~Schmid, ``Orbit determination accuracies using
  satellite-to-satellite tracking,'' \emph{IEEE Transactions on Aerospace and
  Electronic Systems}, no.~6, pp. 834--842, 1978.

\end{thebibliography}

%
% ---- Biography ----

\thebiography
%% This biostyle allows you to insert your photo size 1in X 1.25in
\begin{biographywithpic}
{Ricardo Ferreira}{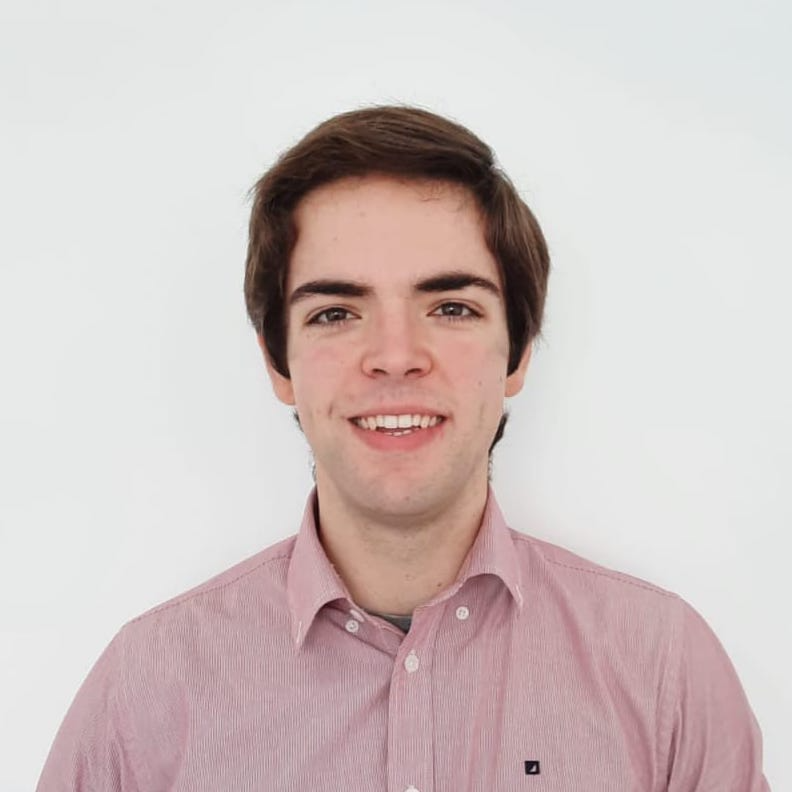}
received his B.Sc. and M.Sc. degree in Computer Science from NOVA School of Science and Technology. He is currently a Ph.D. student in Computer Science at NOVA School of Science and Technology. His research interests include Machine Learning, Optimization and Probability Theory to develop robust solutions for satellite collision avoidance.
\end{biographywithpic}

\begin{biographywithpic}
{Marta Guimarães}{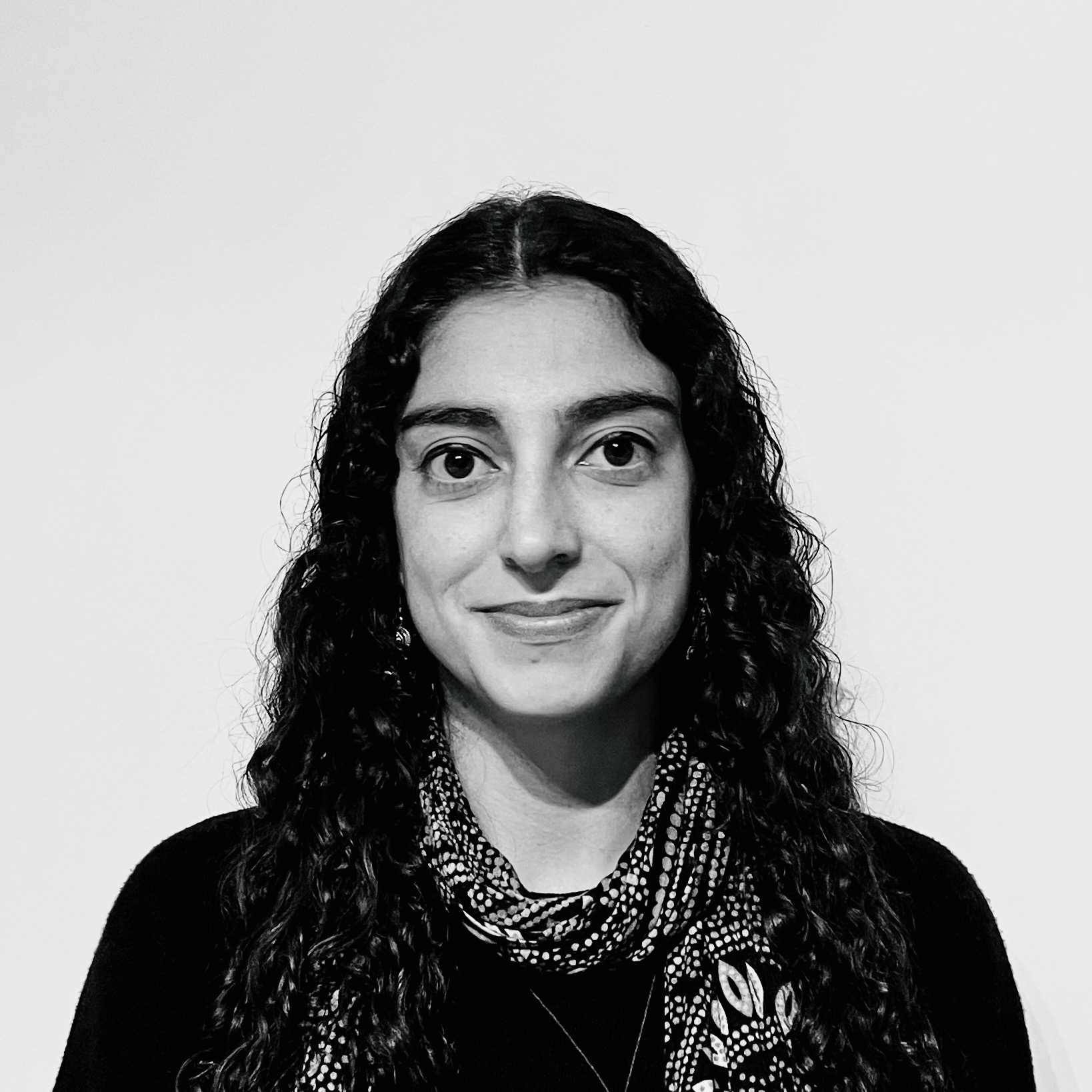}
received her B.Sc. and M.Sc. degrees in Aerospace Engineering from Instituto Superior Técnico. She is currently a Ph.D. student in Computer Science at NOVA School of Science and Technology and works as an AI Researcher at Neuraspace, developing Machine Learning solutions for satellite collision avoidance and space debris mitigation.
Her research interests include Machine Learning and Deep Learning, with a focus on Time Series Forecasting.
\end{biographywithpic}

\begin{biographywithpic}
{Filipa Valdeira}{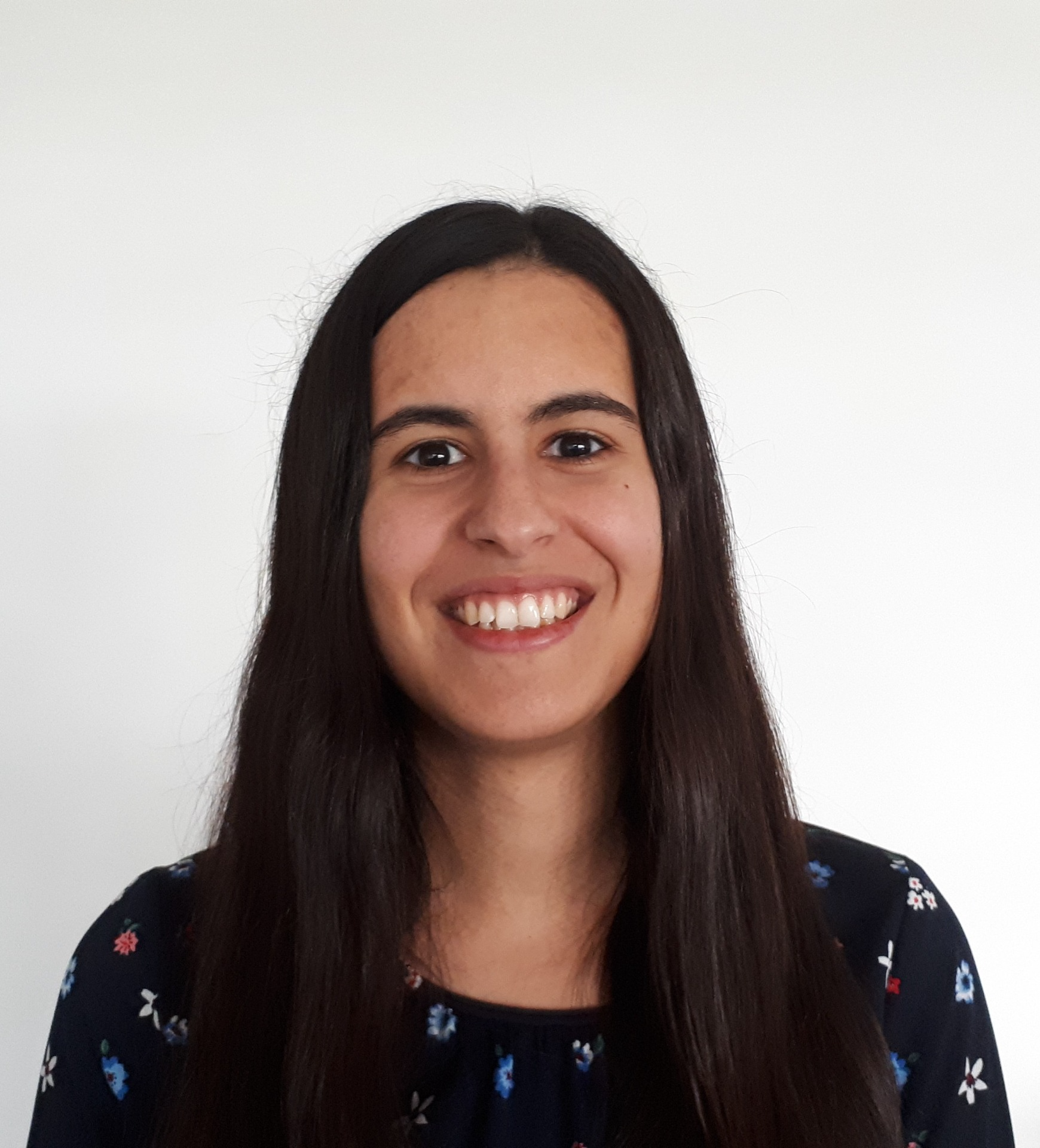}
received her B.Sc. and M.Sc. in Aerospace Engineering from Instituto Superior Técnico in 2018 and a Ph.D. in Mathematical Sciences from the University of Milan in 2022. She is currently a Post-doctoral Researcher at NOVA School of Science and Technology with NOVA LINCS. Her research interests include Machine Learning, Optimization, 3D Shape Modelling and Gaussian Processes.
\end{biographywithpic}

\begin{biographywithpic}
{Cláudia Soares}{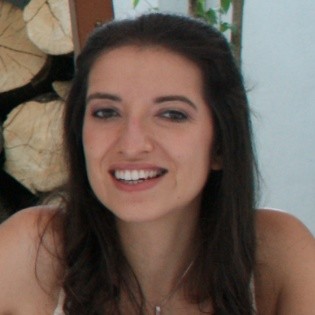}
received a Diploma in modern languages and literature from Nova University of Lisbon, Portugal, as well as a B.Sc., M.Sc., and Ph.D. in Engineering from Instituto Superior Tecnico, Portugal. She is currently an Assistant Professor at NOVA School of Science and Technology, Portugal. Her research focuses on using optimization, physics, and probability theory to develop trustworthy machine learning for various applications, including space.

\end{biographywithpic}

\end{document}